\documentclass[preprint, pre]{revtex4-1}

\usepackage[dvips]{graphicx}

\usepackage{amssymb,amsfonts,amsmath}

\def \>{\rangle} 
\def \<{\langle} 
 
\def\be{\begin{equation}} 
\def\ee{\end{equation}} 
\def\longrightharpoonup{\relbar\joinrel\rightharpoonup}
\def\longleftharpoondown{\leftharpoondown\joinrel\relbar}

\def\longrightleftharpoons{
  \mathop{
    \vcenter{
      \hbox{
      \ooalign{
        \raise1pt\hbox{$\longrightharpoonup\joinrel$}\crcr
	  \lower1pt\hbox{$\longleftharpoondown\joinrel$}
	  }
      }
    }
  }
}

\newcommand \bea {\begin{eqnarray}} 
\newcommand \eea {\end{eqnarray}}

\begin{document}

\title{Statistical mechanics of  transcription-factor binding site discovery using Hidden Markov Models}

\author{Pankaj Mehta}
\email{pankajm@bu.edu}
\affiliation{Dept. of Physics, Boston University, Boston, MA}
\author{David J. Schwab}
\email{dschwab@princeton.edu}
\affiliation{Dept. of Molecular Biology and Lewis-Sigler Institute, Princeton University, Princeton, NJ},
\author{Anirvan M. Sengupta}
\email{anirvans@physics.rutgers.edu}
 \affiliation{BioMAPS and Dept. of Physics, Rutgers University, Piscatway, NJ}

\begin{abstract}
Hidden Markov Models (HMMs) are a commonly used tool for inference of transcription factor (TF) binding sites from DNA sequence data. We exploit the mathematical equivalence between HMMs for TF binding and the ``inverse'' statistical mechanics of hard rods in a one-dimensional disordered potential to investigate learning in HMMs. We derive analytic expressions for the Fisher information, a commonly employed measure of confidence in learned parameters, in the biologically relevant limit where the density of binding sites is low. We then use techniques from statistical mechanics to derive a scaling principle relating the specificity (binding energy) of a TF to the minimum amount of training data necessary to learn it.
\end{abstract}


\maketitle

\section{Introduction}
Biological organisms control the expression of genes using transcription factor (TF) proteins. TFs bind to regulatory DNA segments (6-20bp) called binding sites thereby controlling the expression of nearby genes. An important task in Bioinformatics is identifying TF binding sites from DNA sequence data. This poses a non-trivial pattern recognition problem, and many computational and statistical techniques have been developed towards this goal. The goal of these algorithms is to identify new binding sites starting from a known collection of TF binding sites.  Many different types of algorithms exist including Position  Weight Matrices (PWMs) \cite{Berg88, Stormo98}, biophysics-inspired alogrithms \cite{ Stormo98, Djordjevic03}, Hidden Markov Models (HMMs) \cite{Rajewsky02, Sinha03, Drawid}, and information theoretic algorithms \cite{Kinney07}.

In general, only a limited number of binding sites are known for a given TF.  Thus, any algorithm must build a general classifier based  on limited training data.  This places constraints of the type of algorithms and classifiers that can be used.  The end goal of all models is {\it generalization}--the ability to correctly categorize new sequences that differ from the training set.  This is especially important since the training set is comprised of a small sample fraction of all possible sequences.
Most algorithms create a (often probabilistic) model for whether a particular DNA sequence is a binding site. The model contains a set of parameters, $\theta$, that are fit, or learned, from training data. 

All algorithms exploit the statistical differences between binding sites and background DNA in order to identify new binding sites. Two distinct factors contribute to how well one can learn $\theta$, the size of the training data set and the specificity of the TF under consideration. Many TFs are highly specific. Namely, they bind strongly only to small subset of all possible DNA sequences which are statistically distinct from background DNA. Physically, this means that these TF have large binding energies for certain sequence motifs (binding sites) and low binding energies for random segments of DNA, i.e.  ``background'' DNA. Other TFs are less specific and often exhibit non-specific binding to random DNA sequences.  In this case, the statistical signatures that distinguish binding sites from background DNA are less clear. In general, the more training data one has and the more specific a TF, the easier it is to learn its binding sites.

This raises the natural question: how much data is needed to train an algorithm to learn the binding sites of a TF? In this paper, we explore this question in the context of a widely-used class of bioinformatic methods termed Hidden Markov Models (HMMs).  We exploit the mathematical equivalence between HMMs for TF binding and the ``inverse''  statistical mechanics of hard rods in a one-dimensional disordered potential to derive a scaling principle relating the specificity (binding energy) of a TF to the minimum amount of training data necessary to learn its binding sites. Unlike ordinary statistical mechanics where the goal is to derive statistical properties from a given Hamiltonian, the goal of the `` inverse " problem is to learn the Hamiltonian that most likely gave rise to the observed data.  Thus, we are led to consider a well-studied physics problem \cite{Percus}--the statistical mechanics of a one-dimensional gas of hard rods in an arbitrary external potential--from an entirely new perspective.

The paper is organized as follows. We start by reviewing the mapping between HMMs and the statistical mechanics of hard rods. We then introduce the Fisher Information, a commonly employed measure of confidence in learned parameters, and derive an analytic expression for the Fisher information in the dilute binding site limit. We then use this expression to formulate a simple criteria for how much sample data is needed to learn the binding sites of a TF of a given specificity.

\section{HMMs for binding site discovery}

\begin{figure}
\label{fig:fig1}
\centerline{\includegraphics[width=.6\textwidth]{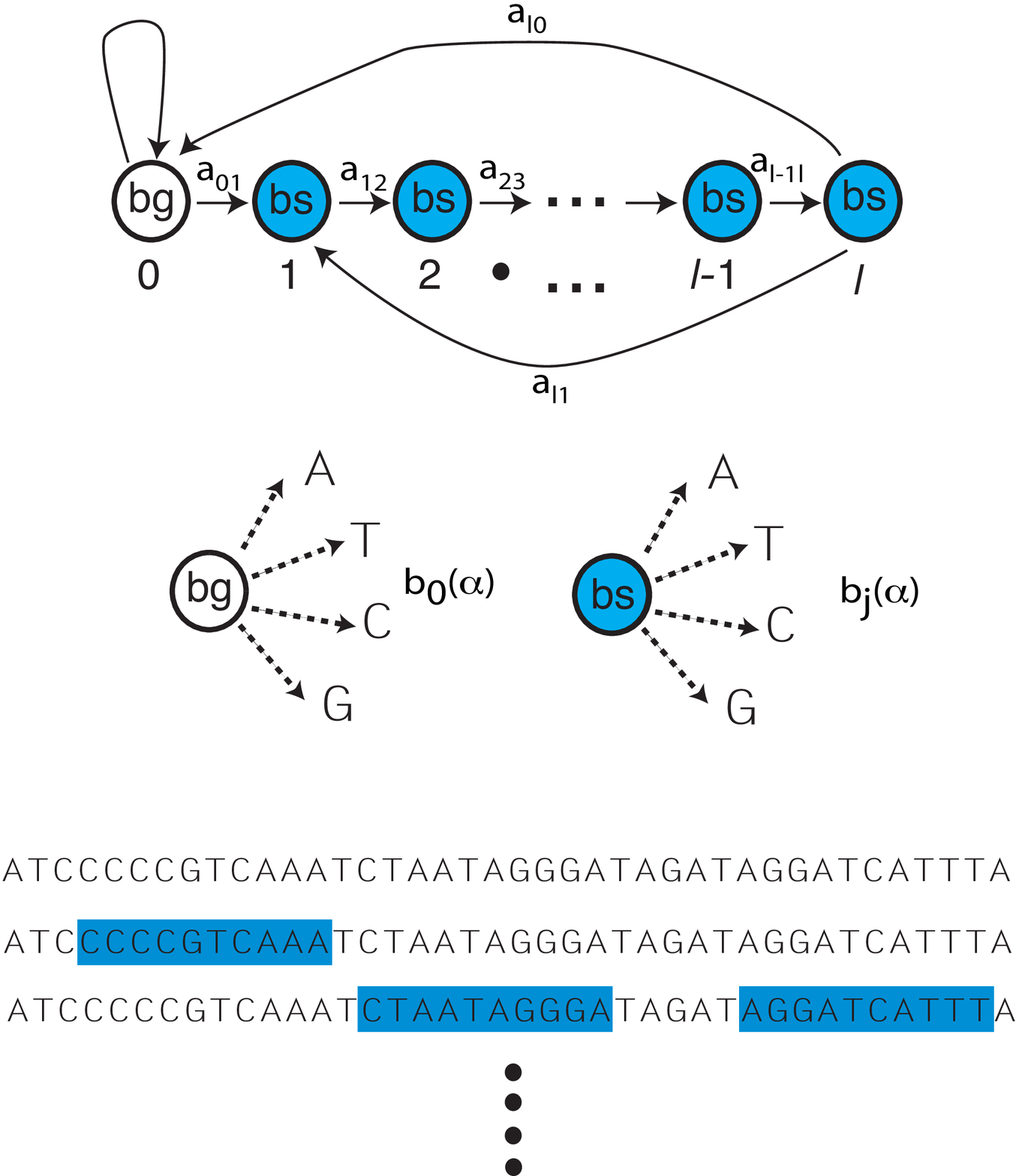}}
\caption{Hidden Markov Model for binding sites of size $l$. (Top) There are $l+1$ hidden states, with state $0$ background DNA and state $j$ corresponding to position $j=1 \ldots l$ in a binding site. The HMM is described by a Markov process with transition probabilities give by $a_{ij}$. (Middle)  Each state $j$ in an HMM is characterized by an observation symbol probability $b_j(k)$, for the probability of seeing symbol $k =A, T,C,G$ in a state $j$ (Bottom) A given sequence of DNA is composed of binding sites and background DNA. }
\end{figure}

HMMs are powerful  tools for analyzing sequential data \cite{Bishop, Rabiner89} that have been adapted to binding site discovery \cite{Sinha03,Drawid}. HMMs model a system as a Markov process on internal states  that are hidden and cannot be observed directly. Instead, the hidden states can only  be inferred indirectly through an observable state-dependent output. In the context of binding site discovery, HMMs serve as generative models for DNA sequences. A DNA sequence is modeled as a mixture of hidden states--background DNA and binding sites--with a hidden state-dependent probability for observing a nucleotide ($A,T,C,G$) at a given location (see Figure \ref{fig:fig1}). 

For concreteness, consider a TF whose binding sites are of length $l$. An HMM for discovering the binding sites can be characterized by four distinct elements (see Fig. 1) \cite{ Rabiner89}: \\
{\bf 1.} $l+1$  hidden states with state $0$ corresponding to background DNA and states $j=1 \ldots l$ corresponding to position $j$ of a binding site.  \\
{\bf 2.}  $4$ observation symbols  corresponding to the four observable nucleotides $\alpha=A,T,C,G$.  \\
{\bf 3.}  The transition probabilities,  $\{a_{ij} \}$ ( $i,j = 0 \ldots l$) between the hidden states which take the particular form shown in Figure \ref{fig:fig1} with  only $a_{j, j+1}$, $a_{l0}$, and $a_{01}$ non-zero. In addition, for simplicity, we assume that binding sites cannot touch (i.e.  $a_{l1}=0$.) The generalization to the case where the last assumption is relaxed is straightforward. 
\\
{\bf 4.} The observation symbols probabilities $\{b_{j}(\alpha)\}$ for seeing a symbol $\alpha=A,C,G, T$ in a hidden state $j$. Often we will rewrite these probabilities in more transparent notation with $p_{\alpha}=b_{0}(\alpha)$, the probability  of seeing base $\alpha$ in the background DNA,  and , $p_{j \alpha}^{(bs)}=b_{j}(\alpha)$ the probability of seeing base $\alpha$ at position $j$ in a binding site. \\
Finally, denote the collection of all parameters of an HMM ($a_{ij}$ and $b_i(\alpha)$)  by the symbol $\theta$.

A DNA sequence of length $L$, ${\cal S}=s_1s_2 \ldots s_L$, is  generated by an HMM  starting in a hidden state $q_1$ as follows. Starting with $i=1$, choose  $s_i$ according to $b_{q_i}(s_1)$ and  then switch to a new  hidden state $q_{i+1}$ using the switching probabilities $a_{q_i q_{i+i}}$ and repeat this process until $i=L$. In this way, one can associate a probability, $p({\cal S}|\theta)$, to each sequence ${\cal S}$, corresponding to the probability of generating $\cal{S}$ using an HMM with parameters $\theta$. The goal of bioinformatic approaches is to learn the parameters $\theta$ from training data and use the result to predict new binding sites. Many specialized algorithms, often termed dynamic programming in the computer science literature, have been developed to this end \cite{Bishop, Rabiner89}.

\subsection{Mapping HMMs to the statistical mechanics of hard rods}

Before discussing the mapping between HMMs and statistical mechanics, we briefly review the physics of a one-dimensional gas of hard rods in a disordered external field \cite{Percus}.  The system consists of hard rods--one-dimensional hard core particles--of length $l$ in a spatially dependent binding energy $E(S_{x_i})$, with $x_i$ the location of the starting site, at a inverse temperature $\beta$, and a fugacity, z (i.e. chemical potential $\mu=\log{z}$). The equilibrium statistical mechanics of the system is determined by the grand canonical partition function obtained by summing over all possible configurations of hard rods obeying the hard-core constraint \cite{Percus}.  In addition, the pressure can be calculated by taking the logarithm of the grand canonical partition function. Since this model is one-dimensional and has only local interactions, many  statistical properties can be calculated exactly using Transfer Matrix techniques. Consequently, variations of this simple hard rod model have been used extensively to model the sequence dependence of nucleosome positioning \cite{Schwab,Morozov}.

We now discuss the mapping between HMMs and a gas of hard rods. We start by showing that the observation symbol probabilities $b_j(\alpha)$ have a natural interpretation as a binding energy. Consider a DNA sequence $S=s_1 \ldots s_l$, with $l$ the length of a binding site. Denote the corresponding hidden state at the $j$-th position of $S$ by $q_j$.  It is helpful to represent this sequence by a $l$ by $4$  matrix  $S_{j \alpha}$ of DNA of length $l$ where $S_{j \alpha}=1$ if  base $s_i=\alpha$ and zero otherwise. Denote the probability of generating $S$ from background DNA as $P(S| \{q_{j}=0, j=1\dots l \}, \theta)=\prod_{j=1}^l b_0 (s_j)$, and the probability of generating the same sequence within a binding site is  $P(S| \{q_j=j, j=1\dots l \}, \theta)= \prod_j b_j (s_j)$. Note that we can rewrite the ratio of these probabilities as 
\be
\frac{P(S| \{q_j=j, j=1\dots l  \}, \theta)}{P(S| \{q_j=0, j=1\dots l \}, \theta)}= \prod_j \frac{b_i(s_j)}{b_0(s_j)} \equiv e^{-E(S)}
\label{equivratio}
\ee 
where we have defined a ``sequence-dependent'' binding energy
\be
E(S)=\epsilon \cdot S = \sum_{\alpha j} \epsilon_{\alpha j} {S}^{ j \alpha}
\label{equivBE2}
\ee
with
\be
\epsilon_{\alpha j} \equiv -\log{\left(\frac{b_j(\alpha)}{b_0(\alpha)}\right)}=-\log{ \left( \frac{p_{j \alpha}^{\rm (bs)}}{p_{\alpha}} \right)}.
\label{equivPWM}
\ee
Notice that the ratio Eq (\ref{equivratio}) is of a Boltzmann form with a  `binding energy'  that can be expressed in terms of a Position Weight Matrix (PWM),  $\epsilon$,  related to the  observation symbol probabilities Eq. (\ref{equivPWM}).

Now consider a sequence ${\cal S}=s_1 s_2 \ldots s_L$ of length $L \gg l$. In this case, the probability of generating the sequence, $P(\cal{S}|\theta)$, is obtained by summing over  all possible hidden state configurations. Notice that  we can uniquely denote a hidden state configuration by specifying the starting positions within the sequence ${\cal S}$ of all the binding sites, $\{ x_{1} \ldots x_{n} \}$.  The hard-rod constraint means that the only allowed configuration are those where $|x_u - x_v| \ge l+1$ for all $u, v$ (the extra factor of 1 arises because $a_{l0}=0$).  Consequently, the probability of generating a sequence $\cal S$ is given by summing over all possible hidden state configurations 
\be
P({\cal S}|\theta) = \sum_n \sum_{x_1 \ldots x_n} P({\cal S}| \{x_1 \ldots x_n \} , \theta) P(\{x_1 \ldots x_n \} | \theta).
\label{equivpartition}
\ee
where  $P(\{ x_1 \ldots x_n \}| \theta)$ is the probability of generating an allowed hidden state configuration, $\{ x_1, \ldots, x_n \}$ and we have factorized the probability using the fact that in a HMM, transition probabilities are independent of the observed output symbol. Furthermore, the ratio of $P(\{ x_1 \ldots x_n \}| \theta)$ to the probability of generating a hidden-state configuration with no binding sites, $P(\emptyset |\theta)$ is just
\be
\frac{P(\{ x_1 \ldots x_n \}| \theta)}{P( \emptyset |\theta)} = z^n=e^{n \mu }
\label{equivfugacity}
\ee
with the `fugacity', $z$, given by
\be
z = \frac{a_{01}}{(1-a_{01})^{l+1}}.
\label{fugacity}
\ee
and $\mu = \log z$ the chemical potential. Combining Eqs.  (\ref{equivratio}), (\ref{equivfugacity}), and  (\ref{equivpartition}) yields
\be
\frac{P(\cal{S}|\theta)}{C(\cal{S}, \theta)}= {\cal Z}(\cal{S}|\theta) 
\ee
with
\be
{\cal Z}({\cal S}|\theta) = \sum_{n=0}^{L/l} \sum_{x_1 \ldots x_n} e^{-\sum_{x_1 \ldots x_n}E({x_i})}z^n
\label{equivPF}
\ee
and
\be
C({\cal S}, \theta)= a_{00}^{L-1} \prod_{i, \alpha} p_{\alpha}^{{\cal S}_{i \alpha}}
\label{defNormalization}
\ee
where  $E({x_i})$ is the binding energy, Eq. (\ref{equivratio}), for a sub-sequence of length $l$ starting at position $x_i$ of $\cal{S}$.

Notice that  ${\cal Z}(\cal{S}|\theta)$ is the grand canonical partition function for a classical fluid of hard rods in an external potential \cite{Percus}. The sequence-dependence PWM  $\epsilon$ acts as an arbitrary external potential, and the  switching rate $a_{01}$ sets the chemical potential for binding. Thus, up to a multiplicative factor  $C(\cal{S}, \theta)$ that is independent of the emission probabilities for binding sites, an  HMM is mathematically equivalent to a thermodynamic model of hard rods.  Importantly, the amount of training data, $L$, plays the role of system size. Furthermore,  the negative log-likelihood, $-\log{P(\cal{S}|\theta)}$ is, up to a factor of $L$ just the pressure of the gas of hard rods \cite{Percus}. In what follows, we exploit the relationship between system size and the quantity of training data to use insights from finite-size scaling to better understand how much data one needs to learn small differences.  The relationship between HMMs and the statistical mechanics of hard rods is summarized in Table \ref{Table}.

\begin{table}[t]
\caption{Relationship between HMMs and the statistical mechanics of hard-rods.}
\begin{tabular}{@{\extracolsep{\fill}}l | c r}
 \hline
 & HMMs & hard-rods  \\
 \hline
\\
  $L$         & Size of training data & System size \\
  
   $S_j^{ \alpha} $       &  Nucleotide sequence & Disorder\\ 

 $b_j(\alpha)$ & Symbol Probability & Binding Energy\\
 
 $ a_{ij}$ & Switching rates & Fugacity \\
 
  $P({\cal S}|\theta)$& Probability  & Partition Function \\

  $\log{P({\cal S}|\theta)}$ &  Log-likelihood & Pressure  \\
  $ [I(\theta)]_{ij}$ &  Fisher Information &  Correlation Functions \\
 &  Dynamic Prog. & Transfer Matrices \\
 &  Expectation Maximization & Variational Methods \\\\
\hline
\end{tabular}
\label{Table}
\end{table}

\subsection{HMMs, Position-Weight Matrices, and cutoffs}

The matrix of parameters,  $\epsilon_{i \alpha}$, defined in equation (\ref{equivPWM}), are often referred to in bioinformatics as the Position Weight Matrix (PWM) \cite{Berg88, Stormo98}. PWMs are the most commonly used bioinformatic method for discovering new binding sites. In PWM-based approaches, sequences, $S$, whose binding energies, $E(S)=\epsilon \cdot S$ are below some arbitrary threshold, , are considered binding sites. This points to a major shortcoming of PWM based methods- namely the inability to learn a threshold directly from data.  A major advantage of HMM models over PWM-only approached  is that HMMs learn both a PWM, $\epsilon$, and a natural ``cutoff"  through the chemical potential $\mu=\log{z}$  \cite{Drawid}. In terms of the corresponding hard-rod model, the probability, $P_{bs}(S)$, for a sequence, $S$, to be a binding site takes the form of a Fermi-function,
\be
P_{bs}(S)= \frac{1}{1+e^{\epsilon \cdot S- \mu}}.
\ee
If one makes the reasonable assumption that a sequence $S$ is a binding site if $P_{bs}(S) >1/2$, we see that $\mu$ serves as a natural cut-off for binding site energies \cite{Drawid}.
Thus, the switching probabilities $a_{ij}$ of the HMM can be interpreted as providing a natural cut-off for binding energies through (\ref{fugacity}). This points to a natural advantage of HMMs over PWM-only approach, namely one learns the threshold binding energy for determining whether a sequence is a binding site self-consistently from the data.  Thus, though in practice binding sites are dilute in the DNA and hard-rod constraints can often be neglected, it is still beneficial to use the full HMM machinery for binding site discovery.

\section{Fisher Information \& learning with finite data }

\subsection{Fisher Information and Error-bars}

In general, learning the parameters of an HMM from training data is a difficult task. Commonly, parameters of an HMM are chosen to maximize the likelihood of  observed data, $\cal S$, through Maximum Likelihood Estimation (MLE), i.e.  parameters are chosen so that 
\be
\hat{\theta}=\arg\max_\theta {\cal L}({\cal S}|\theta) = \arg\max_\theta \log P({\cal S}|\theta).
\label{MLcondition}
\ee

Finding the global maxima is an extremely difficult problem. However, one can often find a local maximum in parameter space, $\hat{\theta}$, using Expectation Maximization algorithms such as Baum-Welch \cite{Baum}. In general, for any finite amount of training data, the learned parameters $\hat{\theta}$ (even if they are a global maxima) will differ from the ``true'' parameters $\theta_T$. The reason for this is that the probabilistic nature of HMMs leads to 'finite size' fluctuations so that the training data may not be representative of the data as a whole. These fluctuations are suppressed asymptotically as the training data size approaches infinity. For this reason, it is useful to have a measure of how well the learned parameters $\hat{\theta}$ describe the data.

\begin{figure}[t]
\includegraphics[width=.8\textwidth]{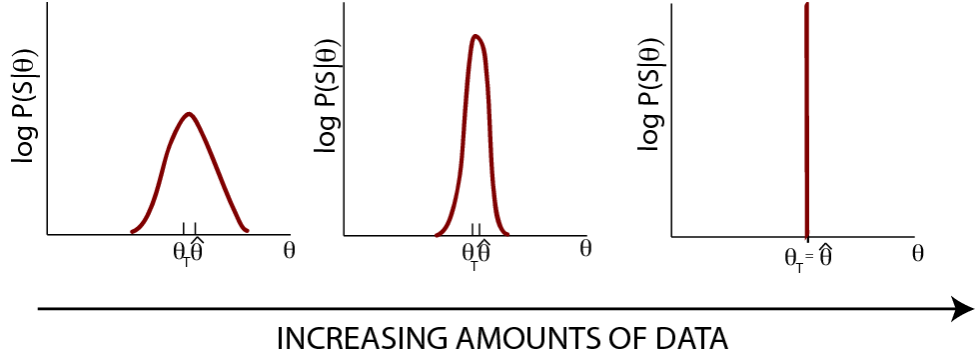}
\caption{The log-likelihood becomes peaked around true parameters with increasing data, analogous to finite-size scaling of pressure times volume (logarithm of the grand canonical partition function) in the corresponding statistical mechanical model.}
\label{fig:fig2}
\end{figure}

In the remainder of the paper, we assume there is enough training data to ensure that we can consider parameters in the neighborhood of the true parameters . The mapping between HMMs and the statistical mechanics of hard rods allows us to gain insight into the relationship between the amount of training data and the confidence in  learned parameters. Recall that  log-likelihood per unit volume,  ${\cal L}({\cal S}|\theta)/L$, is analogous to a pressure and the amount of training data is just the system size. From finite-size scaling in statistical mechanics, we know that as $L \rightarrow \infty $ the log-likelihood/pressure becomes increasingly peaked around its true value (see Fig. \ref{fig:fig2}). In addition, we can approximate the uncertainty we have about parameters by calculating the curvature of the log-likelihood,  $\partial^2_{AB}  {\cal L}({\cal S}| \theta)$, around $\hat{\theta}$ where $\partial_A$ denotes the derivative with respect to the $A$-th parameter.

This intuition can be formalized for MLE using the Cramer-Rao bound which relates the covariance of estimated parameters to the Fisher Information (FI) Matrix, ${\cal I}_{AB}(\theta)$, defined by
\be
[{\cal I}(\theta)]_{AB}= -E_\theta\left[ \partial^2_{ AB} {\cal L}({\cal S}|{\theta}) \right] 
\label{defFisher2}
\ee
where $E_\theta[g({\cal S})]= \sum_{{\cal S}} p({\cal S}|\theta) g({\cal S})$  (\cite{Bishop} and see Appendix). An important property of the Fisher information is that it provides a bound for how well one can estimate the parameters of the likelihood function by placing a lower bound on the covariance of the estimated parameters. The Cramer-Rao bound relates the Fisher information to the expected value of an unbiased estimator,
$ E_\theta[\hat{\theta}({\cal S})]$, and the covariance matrix of the estimator,
\bea
[{\rm Cov}_{\theta}(\hat{\theta})]_{AB} &\equiv& E_\theta[(\hat{\theta}_A({\cal S})-\theta_A)(\hat{\theta}_B({\cal S})-\theta_B)], \nonumber
\label{defCov}
\eea
through the inequality
\be
[{\rm Cov}_{\theta}(\hat{\theta})]\ge  [{{\cal I}(\theta)}]^{-1}.
\label{CRbound}
\ee
For MLE, the Cramer-Rao bound is asymptotically saturated in the limit of infinite data.  Thus, we expect the Fisher Information to be a good approximation for ${\rm Cov}_\theta[\hat{\theta}]$ when the amount of training data is large. In the limit of large data, the pressure, or equivalently ${\cal L}({\cal S}|\theta)$,  ``self-averages''  and we can ignore the expectation value (\ref{defFisher2}).  Thus, to leading order in $L$, one can approximate the covariance matrix as
\be
[{\rm Cov}_{\theta}(\hat{\theta})]_{AB}  \approx [{\cal I}(\theta)]^{-1}_{AB} \approx -\left[ \partial^2_{AB}  {\cal L}({\cal S}|\theta) \right]^{-1},
\label{FIfinal}
\ee
in agreement with intuition from finite size scaling. The previous expression provides a way to put error bars on learned parameters. However, in practice we seldom have access to the ``true'' parameters $\theta$ that generated the observed sequences. Instead, we only know the parameters learned from the training data, $\hat{\theta}$. Thus, one often substitutes  $\hat{\theta}$, our best guess for the parameters $\theta$ in Eq. (\ref{FIfinal}).

\subsection{Fisher Information as correlation functions}

It is worth noting that the expression above, in conjunction with the mapping to the hard-rod model, allows us to calculate error bars directly from data. In particular, we show below that the Fisher information can be interpreted as a correlation function and thus can be calculated using Transfer Matrix techniques.  It is helpful to reframe the discussion above in the language of the statistical mechanics of disordered systems. Recall that  up to a normalization constant, $C({\cal S}, \theta)$, in the corresponding hard-rod model $P({\cal S}|\theta)$ is  the grand canonical partition function, ${\cal L}({\cal S}|\theta)/L$ is the pressure, and the amount of  training data, $L$, is just the size of the statistical mechanical system (see Table 1 of main text). When $L$ is large, we expect that the sequence ${\cal S}$ self-averages and the Fisher Information is related to the second derivative of the log-likelihood of the observed data,
\be
[{\cal I}(\theta)]_{AB} \approx - \partial^2_{AB}  {\cal L}({\cal S}|\theta).
\ee
Thus, aside from the the normalization $C({\cal S}, \theta)$, the Fisher Information can be calculated from the second derivative of the pressure. From the fluctuation dissipation theorem, we conclude that the Fisher information can be expressed in terms of connected correlation functions. In particular, let ${\cal O}_A$ be the operator conjugate to the $A$-the parameter,  $\theta_A$ in the partition function ${\cal Z}_{\cal S}(\theta)$. The Fisher Information then takes the form
\be
[{\cal I}(\theta)]_{AB}=  \< {\cal O}_A {\cal O }_B\>_c  - \partial_{AB} \log{C({\cal S},\theta)}.
\label{Fishercorrfunc}
\ee
where $  \< {\cal O}_A {\cal O }_B\>_c =   \< {\cal O}_A {\cal O }_B\> - \< {\cal O}_A\>\< {\cal O }_B\>$ and
\be
\< {\cal O} \> = \frac{\sum_{n=0}^{L/l} \sum_{x_1 \ldots x_n} e^{-E(S_{x_i})}z^n {\cal O} }{{\cal Z}_{\cal S}(\theta)}
\ee
Note that these correlation functions can be calculated directly from the data using Transfer Matrix techniques without resorting to more complicated methods.

In general, the background statistics of the DNA are known and the parameters one wishes to learn are the switching rates, $a_{ij}$, and symbol observation probabilities, $b_j(\alpha)$. In practice, it is often more convenient to work with the fugacity, $z$, rather than the switching rate (see Table 1).  The operator conjugate to the fugacity is $n$, the number of binding sites. Consequently,  
\be
[{\cal I}(\theta)]_{zz}= \< (n-\<n\> )^2\> + \partial_{zz} \log{C({\cal S},\theta)}.
\ee
Thus, the uncertainty in the switching rates is controlled by the fluctuations in binding site  number,  as is intuitively expected. One can also derive the conjugate operators for  the emission probabilities $b_j(\alpha)$ and/or  the sequence dependent  ``binding energies''  $\epsilon_{j \alpha}$ (see Table 1) via  a straight forward calculation (see calculations in sections below). 

The expression (\ref{Fishercorrfunc}) provides a computationally tractable way to calculate the Fisher Information and, consequently, the covariance matrix $[{ \rm Cov}_{\theta}(\hat{\theta})]_{AB}$. Not only can we learn the maximum likelihood estimate for parameters, we can also put `error bars'  on the MLE. We emphasize that in general, this requires powerful, computationally intensive techniques.  However, by exploiting transfer matrix/ dynamic programming techniques, the correlation functions (\ref{Fishercorrfunc}) can be computed in polynomial time. This result highlights how thinking about HMMs in the language of statistical mechanics can lead to interesting new results.

\section{Analytic expression using a Virial Expansion}

In general, calculating the log-likelihood ${\cal L}({\cal S}, \theta)$ analytically is intractable. However, we can exploit the fact that binding sites are relatively rare in DNA and perform a Virial expansion in the density of binding sites, $\rho$, or in the HMM language, the switching rate  from background to binding site ($a_{01} \ll 1$). This is a good approximation in most cases. For example, for the NF-$\kappa$B TF family, $a_{01}$ was recently found to be of order $10^{-2}-10^{-4}$ \cite{Drawid} Thus, to leading order in $\rho$, we can ignore exclusion effects due to overlap between binding sites and write the partition function of the hard-rod model as 
\be
{\cal Z}_{{\cal S}}(\theta)=  \sum_{n=0}  \prod_{x_1 \ldots x_n} e^{-E(x_i)}z^n \approx \prod_{\sigma=1}^{L} (1+ze^{-E(S^\sigma)}) + \sigma(\rho^2)
\ee
where $E(S^\sigma)$ is the binding energy,  (\ref{equivBE2}), for a hard-rod  bound to a sequence, $S^\sigma$, of length $l$ starting at  position $\sigma$ on the full DNA sequence ${\cal S}$. The corrections due to steric exclusion are higher order in density and thus can be ignored to leading order. Thus, the log-likelihood takes the simple form
\be
{\cal L}({\cal S}|\theta) \approx \sum_{\sigma} \log{(1+ze^{-E(S^\sigma)} )} - \log{C({\cal S}, \theta)}.
\label{VirialLikelihood}
\ee
where $C({\cal S}, \theta)$ is a normalization constant.

Notice the log-likelihood (\ref{VirialLikelihood}) is a sum over the free-energies of single particles in potentials given by the observed DNA. For long sequences where $L \gg 1$, we expect on average $N=La_{01}$ binding sites, and $L-N$ background DNA sequences in the sum. In this case, we expect that the single particle energy self-averages and we can replace the sum by the average value of the single-particle free energy in either background DNA or a binding site. In particular, we expect that
\be
{\cal L}({\cal S}|\theta) \approx N \< \log{(1+ze^{-E(S)} )} \>_{bs}+ (L-N) \< \log{(1+ze^{-E(S)} )} \>_{bg} -\log{C({\cal S}, \theta)}
\ee
where $\< H(S) \>_{bg}$ and $\< H(S) \>_{bs}$ are the expectation value of $H(S)$ for sequences $S$ of length $l$ drawn from the background DNA and binding site distributions, respectively. 

\subsection{Maximum Likelihood  Equations via the Virial expansion}

We now derive the Maximum-Likelihood equations  (MLE) within the Virial expansion to the log-likelihood (\ref{VirialLikelihood}).  Recall from (\ref{MLcondition}) that the Maximum Likelihood estimator is the set of parameters most likely to generate the data. Thus, we can derive MLE by taking the first derivatives of  the log-likelihood and setting the expressions to zero. Consider first  the MLE for the binding energy matrix $\epsilon_{i \alpha}$. Since $C(S, \theta)$ is independent of the binding energy, we focus only on the first term of (\ref{VirialLikelihood}). Define the matrix $S_{i \alpha}$ which is one if position $i$ has base $\alpha$ and zero otherwise. The MLE can be derived by taking the first derivative
\bea
\partial_{\epsilon_{i \alpha} }[\sum_{\sigma} \log{(1+ze^{-E(S^\sigma)} )} + \sum_{i} \lambda_i( \sum_{\alpha} p_{\alpha}e^{-\epsilon_{i \alpha}}-1)] \nonumber  \\
=\partial_{\epsilon_{i \alpha}} [\sum_{\sigma} \log{(1+ze^{-E(S^\sigma)} )} + \sum_{i} \lambda_i( \sum_{\alpha}  p_{\alpha i}^{(bs)}-1)]
\eea
where $\lambda_i$ are Langrange multipliers that ensure proper normalization of probabilities..  Explicitly taking the derivative, using probability conservation, and noticing that $\sum_{\alpha} S_{i \alpha}=1$ gives
\be
\frac{  \sum_\sigma f_{z,\epsilon}(S^\sigma) S_{i \alpha}}{\sum_{\sigma} f_{z, \epsilon}(S^\sigma)}=p_{\alpha}e^{-\epsilon_{i \alpha}} =p_{i \alpha}^{(bs)}
\label{1TFMLE}
\ee
where 
\be
f_{z, \epsilon}(S^\sigma)= \frac{1}{1+z^{-1}e^{E( S^\sigma)}}
\ee
is the Fermi-Dirac distribution function.

We can also derive the MLE corresponding to the fugacity. The fugacity depends explicitly on the normalization constant $C({\cal S}, \theta)$.  Note that in HMMs, $C({\cal S},\theta)= a_{00}^L \prod_{\sigma} p_{\alpha}^{{\cal S}_{i \alpha}}$ and ensures probability conservation.  Since $z=a_{01}/(1-a_{01})^{l+1}$, to leading order in $a_{01}$, naively $\log C({\cal S}, \theta) \sim  L \log(1-z)$. However, choosing this normalization explicitly violates probability conservation in the corresponding HMM because we have truncated the Virial expansion for the log-likelihood at first order and consequently allowed unphysical configurations.  Since deriving the MLEs requires probability conservation, we impose by hand that the normalization has the $z$ dependence,
\be
\log C({\cal S}, \theta) \sim L \log(1+z).
\label{NormMixture}
\ee
With this normalization, the log-likelihood (\ref{VirialLikelihood}) becomes analogous to that for a mixture model where the sequences $S^\sigma$ are drawn from background DNA or binding sites.  With this choice of $C(S, \theta)$ the MLE equations can be calculated in a straight forward manner by taking the derivative of  (\ref{VirialLikelihood}) with respect to $z$  (see Appendix B) to get
\be
\sum_{\sigma} f_{z, \epsilon} (S^\sigma) = \frac{Lz}{1+z}.
\ee

\subsection{Fisher Information via the Virial Expansion}

One can also  derive an analytic expressions for the Fisher information within the Virial expansion. Generally, the background observation probabilities $b_0(\alpha)=p_{\alpha}$ are known and  the HMM parameters, $\theta$, to be learned are the observation symbol probabilities in binding sites, $b_j(\alpha)=p_{j \alpha }^{\rm (bs)}$ and the switching probability $a_{01}$. Technically, it is easier to work with the corresponding parameters of the hard-rod model, the binding energies $\epsilon_{i \alpha}$ and the fugacity, $z$.  Note that probability conservation and (\ref{equivPWM}) imply that only three of the $\epsilon_{i \alpha}$  ($\alpha=A,C,G$) are independent. A straight forward calculation (see Appendix) yields  
\be
 [{\cal I}(\theta)^{-1}]_{i \alpha, j \beta} \approx N \<A_{i\alpha,j\beta} \>_{bs} +(L-N) \<A_{i\alpha,j\beta} \>_{bg}
\label{FIanalyic}
\ee
with 
\be
A_{i\alpha,i\beta}= [f_{z, \epsilon}(S) ]^2 \left[ \delta_{\alpha \beta} S_{i \alpha} S_{i  \beta} + \frac{p_{i \alpha}^{\rm (bs)}p_{i \beta}^{\rm (bs)}}{(p_{i T}^{\rm(bs)})^2} {S}_{i T} { S}_{i T} \right]
\ee
and for $i \neq j$,
\be
A_{i\alpha,( j \neq i) \beta} =-f_{z, \epsilon}( S)(1-f_{z, \epsilon}( S) )
\left[ {S}_{i \alpha} -\frac{p_{i \alpha}^{\rm (bs)}}{p_{i T}^{\rm(bs)}} { S}_{i T} \right] \left[ {S}_{j  \beta} -\frac{p_{j \beta}^{\rm(bs)}}{p_{j T}^{\rm(bs)}} {S}_{j  T}\right] \nonumber
\ee
 where, as above, $f_{z, \epsilon}({ S})$ is the Fermi-Dirac distribution function 
\be
f_{z, \epsilon}({S})=\frac{1}{1+z^{-1}e^{E(S)}}.
\ee
One also has (see Appendix B)
\be
 [{\cal I}(\theta)^{-1}]_{i \alpha, z} \approx N \<C_{i \alpha} \>_{bs} +(L-N) \<C_{i\alpha} \>_{bg},
\label{FIanalytic2}
\ee
with
\be
C_{i \alpha}=\frac{1}{z} f_{z, \epsilon}(S)(1-f_{z, \epsilon}(S))S_{i \alpha},
\ee
and
\be
 [{\cal I}(\theta)^{-1}]_{z, z} \approx  N \<D \>_{bs} +(L-N) \<D \>_{bg},
\ee
with
\be
D= \frac{f_{z, \epsilon}(S)}{z^2}- \frac{1}{(1+z)^2}.
\label{FIanalyic3}
\ee

The expressions (\ref{FIanalyic}),(\ref{FIanalytic2}), and  (\ref{FIanalyic3}) depend only on  $\epsilon_{i \alpha}$ and thus can be used to calculate the expected error in learned parameters as a function of training data using only the Position Weight Matrix  (PWM) of a transcription factor and a rough estimate of the switching probability $a_{01}$ or equivalently the fugacity $z$. The explicit dependence on base $T$ reflects the fact that not all the elements of the PWM are independent.

\section{Scaling relation for learning with finite data}

An important issue in statistical learning is how much data is needed to learn the parameters of a statistical model.  The more statistically similar the binding sites are to background DNA  (i.e the smaller the binding energy of a TF), the more data is required to learn the model parameters. The underlying reason for this is that the probabilistic nature of HMMs means that the training data may not be representative of the data as a whole. Intuitively, it is clear that in order to be able to effectively learn model parameters, the training data set should be large enough to ensure that  ``finite-size'' fluctuations resulting from limited data cannot mask the statistical differences between binding sites and background DNA. To address this question, we must consider PWMs learned from strictly random data. As the size of the training set is increased, the finite-size fluctuations are tamed. Our approach is then, in a sense, complementary to looking for rare, high-scoring sequence alignments which become $\it{more}$ likely as $L$ increases in random data \cite{hwa2}. Of course, estimations based on random data neglect non-trivial structure of real sequences \cite{Tanay}.

\subsection{Maximum Likelihood and Jeffreys priors}
Within the Maximum Likelihood framework, the probability that one learn a ML estimator, $\hat{\theta}$, given that the data is generated by parameters $\theta$, can be approximated by a Gaussian whose width is related to the Fisher information using a Jeffreys prior \cite{Jeffreys46},
\be
P(\hat{\theta}) \propto \sqrt{|{\cal I}(\theta)|}e^{-(\hat{\theta}-\theta_0)[{\cal I}(\theta)]^{-1}(\hat{\theta}-\theta_0)}.
\label{GaussianMLE}
\ee
As expected, the width of the Gaussian is set by the covariance matrix for $\hat{\theta}$, and is related to the second derivative of the log-likelihood through (\ref{FIfinal}). Since the log-likelihood--in analogy with the pressure (times volume) of the corresponding hard-rod gas--is an extensive quantity, an increase in the amount of training data $L$ means a narrower distribution for the learned parameters $\hat{\theta}$ (see Fig. \ref{fig:fig2}). When $L$ is large, the inverse of the Fisher information is well approximated by the Jacobian of the log-likelihood, (\ref{FIfinal}). In general, the Jacobian is a positive semi-definite, symmetric  square matrix of dimension $n$, with $n$ the number of parameters needed to specify the position weight-matrix and fugacity for a single TF. In most cases, $n$ is large and typically ranges from $24-45$, with the exact number equal to three time the length of a binding site. 

Label the $A$-th component of $\theta$ by $\theta_A$. Then, the probability distribution (\ref{GaussianMLE}) can also be used to derive a distribution for the Mahalanobis distance \cite{mahalanobis2050generalized}
\be
\hat{r}^2 = - \sum_{A, B}   [\hat{\theta}_A -\theta_{0A}] \frac{\partial^2 {\cal L}({\cal S}|\theta)}{\partial \theta_A \partial \theta_B}|_{\theta_0}[\hat{\theta}_B -\theta_{0B}].
\ee
The Mahalanobis distance is a scale-invariant measure of how far the learned parameters $\hat{\theta}$ are from the true parameters $\theta$. Intuitively, it measures distances in units of standard deviations. Furthermore, the Mahalanobis distance scales linearly with the amount of data/system size $L$ since it is proportional to log-likelihood ${\cal L}({\cal S}|\theta_0)$ (i.e. pressure times volume).   By changing variable to the eigenvectors of the Jacobian, normalizing by the eigenvalues, and intergrating out angular variables, one can show that (\ref{GaussianMLE}) yields the following distribution for $\hat{r}$,
\be
P(\hat{r}) \propto \, {\hat r}^{n-1} e^{-\hat{r}^2} =  e^{-\hat{r}^2+(n-1)\log{\hat{r}}}. 
\ee
When $n$ is large, we can perform a saddle-point approximation for $r$ around its maximum value,
\be
\hat{r}_*=\sqrt{(n-1)/2}.
\ee 
Writing $\hat{r} =\hat{r_*}+\delta \hat{r}$, one has
\bea
P(\delta \hat{r}) &\approx & e^{-(n-1)/2+\log{(n-1)/2}} e^{-\delta \hat{r}^2}.
\label{MDdistribution}
\eea
Thus, for large $n$, almost in all cases the learned parameters $\hat{\theta}$ will be peaked sharply around a distance, $\hat{r}_*^2=(n-1)/2$, with a width of order $1$. This result is a general property of large-dimensional Gaussians and will be used below.

\subsection{Scaling Relation for learning with finite data}
We now formulate a simple criteria for when there is enough data to learn the binding sites of a TF characterized by a PWM $\epsilon$. We take as our null hypothesis that the data was generated entirely from background DNA (i.e the true parameters are $\epsilon_0=0$ and $z_0=z$)  and require enough data so that the probability of learning $\hat{\epsilon}=\epsilon$ be negligible. In other words, we want to make sure that there is enough data so that there is almost no chance of learning $\hat{\epsilon}=\epsilon$ for data generated entirely from background DNA, $\epsilon_0=0$. From (\ref{MDdistribution}), we know that for large $n$, with probability almost 1 due to finite size fluctuations, any learned $\hat{\epsilon}$ will lie a Mahalanobis distance, $r^2(\hat{\epsilon},z)= (n-1)/2$ away from the true parameters. Thus, we require enough data so that 
\be
r^2(\epsilon,z)\equiv L \tilde{r}^2(\epsilon, z) \ge (n-1)/2,
\label{GeneralScaling}
\ee
with $ \tilde{r}^2(\epsilon, z)$ defined by the first equality.
An explicit calculation of the left hand side of  (\ref{GeneralScaling}) yields  (see Appendix) 
\be
\tilde{r}^2(z, \epsilon)= \frac{z^2}{(1+z)^2}\left[ \sum_i \overline{\epsilon_i^2} + \frac{ \overline{\epsilon_i}^2}{p_T} \right] 
\label{MDcalculation}
\ee
where we have defined
\be
\overline{\epsilon_i^\gamma}= \sum_{\alpha=A,C,G} p_{i \alpha} \epsilon_{i \alpha}^\gamma, \hspace{0.5in} \gamma=1,2.
\ee
Together, (\ref{GeneralScaling}) and (\ref{MDcalculation}) define a criteria for how much data is needed to learn the binding sites of a TF with PWM (binding energy), $\epsilon$, whose binding sites occur in background DNA with a fugacity $z$.Notice that (\ref{MDcalculation}) contains terms that scale as the square of the energy difference, indicating that it is much easier to learn binding sites with  a few large differences than many small differences.

\section{Discussion}

In this paper, we exploited the mathematical equivalence between HMM models for TF binding and the ``inverse'' statistical mechanics of hard rods in a one-dimensional disordered potential to investigate learning in HMMs. This allowed us to derive a scaling principle relating the specificity (binding energy) of a TF to the minimum amount of training data necessary to learn its binding sites. Thus, we were led to consider a well-studies physics problem \cite{Percus}--the statistical mechanics of a one-dimensional gas of hard rods in an arbitrary external potential--from an entirely new perspective. 

In this paper, we assumed that there was enough data so that we could focus on the neighborhood of a single maximum in the Maximum Likelihood problem. However, in principle, for very small amounts of data, the parameter landscape has the potential to be glassy and possess many local minima of about equal likelihood.  However in our experience, this does not seem to be the case in practice for most TFs. In the future, it will be interesting to investigate the parameter landscape of HMMs in greater detail to understand when they exhibit glassy behavior.

The work presented here is part of a larger series of works that seeks to use methods from ``inverse statistical mechanics" to study biological phenomenon \cite{Mora, Bialek, Rivoire, Hwa}. Inverse statistical mechanics inverts the usual logic of statistical mechanics where one starts with a microscopic Hamiltonian and calculates statistical properties such as correlation functions. In the inverse problem, the goal is to start from observed correlations and find the Hamiltonian from which they were most likely generated. In the context of binding site discovery, considering the inverse statistical problem allows us to ask and answer new and interesting questions about how much data one needs to learn the binding sites of a TF. In particular, it allows us to calculate error bars for learned parameters directly from data and derive a simple scaling relation between the amount of training data and the specificity of TF encoded in its PWM.

Our understanding of how the size of training data affects our ability to learn the parameters in inverse statistical mechanics is still in its infancy. It will be interesting to see if the analogy between finite-size scaling in the thermodynamics of disordered systems and learning in inverse statistical mechanics holds in other systems, or if it is particular to the problem considered here. More generally, it will be interesting to see in methods from physics and statistical mechanics yield new insights about large data sets now being generated in biology.

\begin{acknowledgments}
We would like to thank Amor Drawid and the Princeton Biophysics Theory group for useful discussion. This work was partially supported by NIH Grants K25GM086909 (to PM) and R01HG03470 (to AMS). DS was partially supported by DARPA grant HR0011-05-1-0057 and NSF grant PHY-0957573. PM would also like to thank the Aspen Center for Physics where part of this work was completed.

\end{acknowledgments}

\bibliography{refs-3}   

\appendix
\section{Covariance Matrix and Fisher Information}

The Fisher information is a commonly employed measure of how well one learns the parameters,  $\theta$, of a probabilistic model from training data, ${\cal S}$. In our context, ${\cal S}$ is the observed DNA sequence and $\theta$ are the parameters of the HMM for generating DNA sequences. The Fisher information matrix, ${\cal I}_{AB}(\theta)$, is given in terms of the log-likelihood, ${\cal L}({\cal S}|\theta)= \log{p({\cal S}|\theta)}$, by 
\bea
[{\cal I}(\theta)]_{AB}^{-1}& \equiv & E_\theta \left[ \partial_A {\cal L}({\cal S}|{\theta}) \partial_B{\cal L} ({\cal S}|{\theta}) \right] \nonumber \\
&=&  \sum_{{\cal S}} p({\cal S}|\theta) \partial_A {\cal L}({\cal S}|{\theta}) \partial_B {\cal L}({\cal S}|{\theta}). 
\label{defFisher}
\eea
where $E_\theta$ denotes the expectation value averaged over different data sets generated using the parameters $\theta$ and $\partial_A$ denotes the partial derivative with respect to the $A$-th component of  $\theta$ .

The Fisher information can also be expressed as a second derivative of the log-likelihood function. This follows from differentiating both sides of the equation
\be
\sum_{{\cal S}} e^{ {\cal L}({\cal S}|{\theta}) }=1 
\ee
with respect to ${\theta}_A$ and ${\theta}_B$ which yields the expression
\be
\sum_{{\cal S}} e^{{\cal L}({\cal S}|{\theta})} \partial_A {\cal L}({\cal S}|{\theta}) \partial_B {\cal L}({\cal S}|{\theta})  + \sum_{{\cal S}} 
e^{{\cal L}({\cal S}|{\theta})} \partial^2_{AB} {\cal L}({\cal S}|{\theta})=0.
\ee
Comparing with  (\ref{defFisher}), we see that the  Fisher information can also be expressed as 
\be
[{\cal I}(\theta)]_{AB}^{-1}= -E_\theta \left[ \partial^2_{AB} {\cal L}({\cal S}|{\theta}) \right] 
= -\sum_{{\cal S}} p({\cal S} |{\theta}) \partial^2_{A B} {\cal L}({\cal S}|{\theta}).
\label{defFisher2}
\ee

An important property of the Fisher information is that it provides a bound for how well one can estimate the parameters of the likelihood function. As discussed in the main text, the parameters of a HMM can be estimated from an observed sequence, ${\cal S}$, using a Maximum Likelihood Estimator (MLE), ${\hat{\theta}}({\cal S})$, defined as
\be
\hat{\theta}({\cal S}) \equiv  \arg\max_{{\theta}} {\cal L}({\cal S}|{\theta}).
\label{defMLE2}
\ee
The Cramer-Rao bound relates the Fisher Information to the expected value of the estimator 
\be
 E_\theta[\hat{\theta}({\cal S})] = \sum_{{\cal S}} \hat{\theta}({\cal S})p({\cal S}|{\theta}),
\ee
and the covariance matrix of the estimator,
\bea
[{\rm Cov}_{\theta}(\hat{\theta})]_{AB} &\equiv& E_\theta[(\hat{\theta}_A({\cal S})-\theta_A)(\hat{\theta}_B({\cal S})-\theta_B)]. \nonumber
\label{defCov}
\eea
For a multidimensional estimator,  the Cramer-Rao bound is given by
\be
[{\rm Cov}_{\theta}(\hat{\theta})]_{AB} \ge  \sum_{C,D} \partial_B E_\theta(\hat{\theta}_C)[{\cal I}(\theta)]_{CD}\partial_B E_\theta(\hat{\theta}_D).
 \ee
To gain intuition, it is worth considering the special case where the estimator is unbiased, $E_S[\hat{\theta}(S)]=\theta$, in which case the Cramer-Rao bound simply reads
\be
[{\rm Cov}_{\theta}(\hat{\theta})]_{AB} \ge  [{{\cal I}(\theta)}^{-1}]_{AB}. 
\ee
Thus, the Fisher information gives a fundamental bound on how well one can learn the parameters of our HMM. 

For MLEs, the Cramer-Rao bound is asymptotically saturated in the limit of infinite data.  Thus, we expect the Fisher Information to be a good approximation for ${\rm Cov}_\theta[\hat{\theta}]$ when the length, $L$, of the  DNA sequences, $S$, from which we learn parameters is long. In this case,
\bea
[{\rm Cov}_{\theta}(\hat{\theta})]_{AB} &\approx& [{\cal I}(\theta)]^{-1}
\eea
The previous expressions provide a way to put error bars on learned parameters. However, in practice we never have access to the ``true'' parameters $\theta$ that generated the observed sequences. Instead, we only know the parameters learned from the training data, $\hat{\theta}$. Thus, we make the additional approximation
 \bea
[{ \rm Cov}_{\theta}(\hat{\theta})]_{AB} &\approx& [{\cal I}(\theta)]^{-1} \approx [{\cal I}(\hat{\theta})]^{-1}. \label{Covfinal}
\eea

\section{Calculation of Fisher Information using a Virial Expansion}

\subsection{PWM dependent Elements}

We now calculate the Fisher information for a HMM for binding sites from a single binding site distribution using the Virial expansion. We are interested in the Fisher information for the parameters $\epsilon_{i \alpha}$ (the energies in the corresponding Position Weight Matrix).  An important  complication  is that  not all the $\epsilon_{i \alpha}$ are independent. In particular, we have
\be
\sum_{\alpha} p_{\alpha} e^{- \epsilon_{i \alpha}}= \sum_{\alpha} p_{\alpha i}^{(bs)}=1
\ee 
Thus, there are only three independent parameters at each position in the binding site. Let us choose $\epsilon_{i T}$ to depend on the other three energies. Rearranging the equation above, one has that
\be
\epsilon_{i T}=  -\log{\left(\frac{1-\sum_{\alpha \neq T} p_{\alpha} e^{- \epsilon_{\alpha i}}}{p_{T }}\right)} \equiv  g_ {\epsilon_{i T}}
\ee 
Taking the first derivative of (\ref{VirialLikelihood}) with respect to $\epsilon_{i \alpha}$ with $\alpha \neq T$ yields
\be
\frac{\partial {\cal L}({\cal S}|\theta) }{\partial \epsilon_{i \alpha}} = - \sum_{\sigma} f_{z,  \epsilon} (S^\sigma) \left[ S_{i\alpha}^\sigma+ \frac{ \partial g_{\epsilon_{i T}}}{\partial \epsilon_{i \alpha} } S^\sigma_{iT} \right]
\ee
Taking the second derivative yields
\be
\frac{\partial^2 {\cal L}({\cal S}|\theta) }{\partial \epsilon_{i \alpha} \partial  \epsilon_{j \beta}} = \sum_{\sigma} f_{z, \epsilon}(S^\sigma)(1-f_{z, \epsilon}(S^\sigma) )
\left[ S_{i\alpha}^\sigma+ \frac{ \partial g_{ \epsilon_{i T}}}{\partial \epsilon_{i \alpha} } S^\sigma_{iT} \right]\left[ S_{j \beta}^\sigma+ \frac{ \partial g_{ \epsilon_{j T}}}{\partial \epsilon_{j \beta}}S^\sigma_{jT} \right]- \delta_{i j} S_{iT}^\sigma f_{z, \epsilon}(S^\sigma) \frac{\partial^2 g_{ \epsilon_{i T}}}{\partial \epsilon_{i \alpha} \partial \epsilon_{i \beta} }
\label{SDLL1}
\ee

We can simplify the expressions further by noting
\bea
\frac{ \partial g_{\epsilon_{i T}}}{\partial \epsilon_{i \alpha} }&=& = -\frac{p_{i \alpha}^{(bs)}}{p_{i T}^{(bs)} }\\
 \frac{\partial^2 g_{ \epsilon_{i T}}}{\partial \epsilon_{i \alpha} \epsilon_{i \beta} }&=& \left[ \delta_{\alpha \beta} \frac{p_{i \alpha}^{(bs)}}{p_{i T}^{(bs)} }+  \frac{p_{i \alpha}^{(bs)}p_{i \beta}^{(bs)}}{(p_{i T}^{(bs)})^2} \right]
 \eea
Plugging in these expressions into (\ref{SDLL1}) yields
\bea
\frac{\partial^2 {\cal L}({\cal S}|\theta) }{\partial \epsilon_{i \alpha} \partial \epsilon_{j \beta}}& =& \sum_{\sigma } f_{z, \epsilon}(S^\sigma)(1-f_{z, \epsilon}(S^\sigma) )
\left[ S_{i\alpha}^\sigma -\frac{p_{i \alpha}^{(bs)}}{p_{i T}^{(bs)}} S_{i T}^\sigma \right] \left[ S_{j \beta}^\sigma -\frac{p_{j \beta}^{(bs)}}{p_{j T}^{(bs)}} S_{j T}^\sigma \right] \nonumber \\
&-& \delta_{ij} S_{i T}^\sigma  f_{z,  \epsilon}(S^\sigma) \left[ \delta_{\alpha \beta} \frac{p_{i \alpha}^{(bs)}}{p_{i T}^{(bs)} }+  \frac{p_{i \alpha}^{(bs)}p_{i \beta}^{(bs)}}{(p_{i T}^{(bs)})^2} \right]
\label{GeneralFIequation}
\eea
The Fisher information is obtained in the usual way from
\be
[{\cal I}(\theta)]_{ \epsilon_{i \alpha} \epsilon_{j \beta}}^{-1}= -\frac{\partial^2 {\cal L}({\cal S}|\theta) }{\partial \epsilon_{i \alpha} \partial \epsilon_{j \beta}}
\label{FI2D}
\ee

\subsubsection{Simplified Equations for  $i=j$}

When $i=j$, we can simplify the equations above using the ML equations (\ref{1TFMLE}) and noting that $S_{i \alpha} S_{i \beta}= \delta_{\alpha \beta} S_{i \alpha}$ and $S_{i \alpha}S_{i T}=0$.Using the expressions above yields
\be
\frac{\partial^2 {\cal L}({\cal S}|\theta) }{\partial \epsilon_{i \alpha} \partial \epsilon_{i \beta}}= 
\sum_{\sigma} f_{z, \epsilon}(S^\sigma)(1-f_{z, \epsilon}(S^\sigma) )\left[S_{i \alpha} \delta_{\alpha \beta}+ \frac{p_{i \alpha}^{(bs)}p_{i \beta}^{(bs)}}{(p_{i T}^{(bs)})^2} S_{iT}^\sigma \right]
- S_{i T}^\sigma  f_{z,  \epsilon}(S^\sigma) \left[ \delta_{\alpha \beta} \frac{p_{i \alpha}^{(bs)}}{p_{i T}^{(bs)} }+  \frac{p_{i \alpha}^{(bs)}p_{i \beta}^{(bs)}}{(p_{i T}^{(bs)})^2} \right]
\ee
From the MLE (\ref{1TFMLE}), we know that 
\be
\sum_{\sigma}  f_{z,  \epsilon}(S^\sigma) S_{iT}^\sigma \frac{p_{i \alpha}^{(bs)}}{p_{i T}^{(bs)} } = \left[\sum_{\sigma }  f_{z, \epsilon}(S^\sigma) S_{iT}^\sigma \right]\left[\sum_{\sigma^\prime }  f_{z,  \epsilon}(S^{\sigma^\prime}) S_{i\alpha}^{\sigma^\prime} \right] /\left[\sum_{\sigma^{\prime \prime} \in BS}  f_{z,\epsilon}(S^{\sigma^{\prime \prime}}) S_{iA}^{\sigma^{\prime \prime}} \right] =\sum_{\sigma}  f_{z,\epsilon}(S^\sigma) S_{i\alpha}^\sigma
\ee
Plugging this into the equations above yields
\be
 \frac{\partial^2 {\cal L}({\cal S}|\theta) }{\partial \epsilon_{i \alpha} \partial \epsilon_{i \beta}}= 
-\sum_{\sigma} [ f_{z, \epsilon}(S^{\sigma^\prime}) ]^2 \left[ \delta_{\alpha \beta} S_{i \alpha}^\sigma S_{i \beta}^\sigma + \frac{p_{i \alpha}^{(bs)}p_{i \beta}^{(bs)}}{(p_{i T}^{(bs)})^2} S_{iT}^\sigma S_{iT}^\sigma \right]=-\sum_{\sigma}A_{ii}(S^\sigma)
\label{FIfor1TF}
\ee
This is the operator $A_{ii}$ in the main text.

\subsubsection{Simplified Equations for  $i \neq j$}

In this case, we know that 
\be
\frac{\partial^2 {\cal L}({\cal S}|\theta) }{\Delta \epsilon_{i \alpha} \Delta \epsilon_{j \beta}} = \sum_{\sigma} f_{z, \epsilon}(S^\sigma)(1-f_{z, \epsilon}(S^\sigma) )
\left[ S_{i\alpha}^\sigma -\frac{p_{i \alpha}^{(bs)}}{p_{i T}^{(bs)}} S_{i T}^\sigma \right] \left[ S_{j \beta}^\sigma -\frac{p_{j \beta}^{(bs)}}{p_{j T}^{(bs)}} S_{j T}^\sigma \right]=- \sum_{\sigma}A_{i\alpha,j \beta}(S^\sigma)
\ee
This is the operator $A_{ij}$ in the main text.

\subsection{Fugacity dependent Elements}

We start by calculating the elements ${\cal I}_{i \alpha,z}$. As before, within the virial expansion
\be
{\cal L}({\cal S}|\theta) \approx \sum_{\sigma} \log{(1+ze^{-\epsilon \cdot \sigma})}-L\log{(1+z)}.
\ee
Thus, we have 
\be
\frac{ \partial {\cal L}}{\partial z}= \sum_{\sigma} \frac{1}{z} f_{z,\epsilon, z}(S^\sigma)-L\frac{1}{(1+z)}= \sum_{\sigma} \frac{e^{-\epsilon \cdot S^\sigma}}{1+ze^{-\epsilon \cdot S^\sigma}}- L\frac{1}{(1+z)}.
\label{derzLL}
\ee
Taking the second derivate with respect to $\epsilon_{i \alpha}$ yields
\be
\frac{ \partial^2 {\cal L}}{\partial \epsilon_{i \alpha} z}= -\sum_{\sigma} \frac{\left(S_{i \alpha}-\frac{p_{i \alpha} S_{iT}^\sigma}{p_{iT}} \right) e^{-\epsilon \cdot S^\sigma}}{(1+ze^{-\epsilon \cdot S^\sigma})^2}=-\sum_{\sigma} \frac{1}{z} f_{z, \epsilon}(S^\sigma)(1-f_{z, \epsilon}(S^\sigma))\left(S_{i \alpha}-\frac{p_{i \alpha} S_{iT}^\sigma}{p_{iT}} \right) 
\label{Matrix1}
\ee
Furthermore, one has
\be
\frac{ \partial^2 {\cal L}}{\partial z^2}= -\sum_{\sigma} \frac{1}{z^2} [f_{z, \epsilon}(S^\sigma)]^2 + L\frac{1}{(1+z)^2}.
\label{Matrix2}
\ee

\subsection{Relating expressions to those in main text}

The equations in the main text follow by noting that the sum over $\sigma$ can be replaced by a sum over expectation value over sequences $S^\sigma$ drawn from the binding site distribution  and background DNA. For an arbitrary function, $H(S)$, of a sequence $S$ of length $l$,
\bea
\sum_{\sigma} H(S^\sigma) &=& \sum_{\sigma \in BS} H(S^\sigma)+ \sum_{\sigma \in BG} H(S^\sigma)\\
&\approx& N \< H(S) \>_{bs} + (L-N) \<H(S) \>_{bg},
\label{selfaveragesum}
\eea
with $N$ the expected number of binding sites in a sequence of length $L$,  and where $\< H(S) \>_{bg}$ and $\< H(S) \>_{bs}$ are the expectation value of $H(S)$ for sequences $S$ of length $l$ drawn from the background DNA and binding site distributions, respectively. Combining (\ref{selfaveragesum}) and  (\ref{FIfinal}) with the expressions above yields the equations in the main text.

\section{Derivation of the scaling relationship}

To derive the scaling relationship, we must calculate the quantity
\be
r^2(z,\epsilon)=-\sum_{ij \alpha \beta} \epsilon_{i \alpha} \frac{\partial^2 {\cal L}({\cal S}|\theta) }{\partial \epsilon_{i \alpha} \partial  \epsilon_{j \beta}} \epsilon_{j \beta} -\sum_{i \alpha } \epsilon_{i \alpha} \frac{\partial^2 {\cal L}({\cal S}|\theta) }{\partial \epsilon_{i \alpha} \partial z}z - \frac{ \partial^2 {\cal L}}{\partial^2 z}z^2,
\ee
where all the second derivatives are evaluated at $\epsilon=0$. Plugging  (\ref{GeneralFIequation}), (\ref{Matrix1}), and (\ref{Matrix2}) into the expression above, one has
\bea
r^2(z,\epsilon)&=& \sum_{\sigma, i}f_{z, \epsilon=0}(S^\sigma)\frac{S_{i T}^\sigma}{p_{T}} \left[ \overline{\epsilon_i^2} + \frac{\overline{\epsilon_i}^2}{p_T} \right]  \nonumber \\
&-& \sum_{\sigma, i, j}f_{z,\epsilon=0}(S^\sigma)(1-f_{z,\epsilon=0}(S^\sigma))\left[\sum_\alpha \epsilon_{i \alpha}S_{i \alpha}^\sigma -\frac{\overline{\epsilon_i} S_{i T}^\sigma}{p_T}\right]\left[\sum_\beta \epsilon_{j \beta}S_{i \alpha}^\sigma -\frac{\overline{\epsilon_j} S_{j T}^\sigma}{p_T}\right] \nonumber \\
&+&\sum_{\sigma, i, \alpha} f_{z,\epsilon=0}(S^\sigma)(1-f_{z,\epsilon=0}(S^\sigma))\frac{\epsilon_{i \alpha}}{z}\left(S_{i \alpha}-\frac{p_{i \alpha} S_{iT}^\sigma}{p_{iT}} \right)\\ \nonumber
&-& \sum_{\sigma} [f_{z, \epsilon=0}(S^\sigma)]^2 + L\frac{z^2}{(1+z)^2}, \nonumber
\label{tempscaling}
\eea
where we have defined
\be
\overline{\epsilon_i^\gamma}= \sum_{\alpha=A,C,G} p_{i \alpha} \epsilon_{i \alpha}^\gamma,
\ee
with $\gamma=1,2$ and used the fact that $p_{i T}=p_T$ when $\epsilon=0$. When $L$ is large, we can replace the sum over $\sigma$ by an expectation value in background DNA,
\be
\frac{1}{L} \sum_{\sigma} \rightarrow  \<\, \>.
\ee
Furthermore,
\bea
\<S_{iT} \>&=&p_{iT} \nonumber \\
\<S_{i \alpha} S_{j \beta} \> &=& p_{i \alpha}p_{j \beta} (1-\delta_{ij}) + p_{i \alpha} \delta_{ij}\delta_{\alpha \beta} \nonumber \\
\<S_{i \alpha} S_{j T}\>&=&p_{i \alpha} p_{j T} (1-\delta_{ij}) \nonumber \\
\<S_{i T}S_{j T}\>&=& p_{iT}p_{jT}(1-\delta_{ij})+p_{iT} \delta_{ij}.
\eea
Plugging these expressions into (\ref{tempscaling}), noting that the third term averages to zero, and simplifying yields
\be
r^2(z, \epsilon)=\frac{Lz^2}{(1+z)^2}\left[ \sum_i \overline{\epsilon_i^2} + \frac{ \overline{\epsilon_i}^2}{p_T} \right] 
\ee
Finally, it is often helpful to define a rescaled version of $r^2(z, \epsilon)$ that makes the dependence of $L$ explicit,
\be
\tilde{r}^2(z, \epsilon) \equiv \frac{r^2(\epsilon,z)}{L}
\ee

\end{document}